\documentclass[aps,superscriptaddress,preprint,showpacs,nofootinbib,floatfix]{revtex4}
\usepackage{bm}
\usepackage{dcolumn}
\usepackage{epsfig}
\usepackage{pstricks}
\begin{document}

\title{Relativistic description of the $np\to\eta d$
reaction near threshold}

\author{H. Garcilazo}
  \affiliation{Escuela Superior de F\' \i sica y Matem\'aticas
Instituto Polit\'ecnico Nacional, Edificio 9,
07738 M\'exico D.F., Mexico}

\author{M. T. Pe\~na }
  \affiliation{
    Instituto Superior T\'ecnico, 
Centro de F\'{\i}sica das Interac\c c\~oes Fundamentais,
    and Department of Physics, 
   Av. Rovisco Pais, 1049-001 Lisboa, Portugal}

\date{\today}
\begin{abstract}
Relativistic effects in a three-body calculation
of the $np\to\eta d$  process are considered.
Relativistic effects on the range and strength 
of the pion exchange  
contribution to the reaction mechanism
may be large, but boost effects of the two-body interactions
are negligible.
The relativistic
calculation confirms previous non-relativistic results,
showing that the shape of the cross section
near threshold is essentially determined 
by the $\eta d$ final-state interaction alone. 
As for the region away from threshold, the relativistic pion
exchange contribution is seen to
dominate the other mechanisms of the reaction. It turns out that,
within the relativistic
reaction model, the $np\to\eta d$ experimental data excludes
large values for the real part of the $\eta N$ scattering length.
\end{abstract}

\pacs{21.30.Fe, 21.45.+v,25.10.+s,11.80.Jy}
\keywords{}
\maketitle

\section{Introduction}
We investigate relativistic effects in the reaction $np\to\eta d$,
in a calculation that considers the
$\eta d$ final state three-body distortion.
The inclusion of this interaction is crucial for the interpretation of the observed behavior of the cross-section
at threshold \cite{CAL1,CAL2}.
In previous works \cite{GAR1,GAR2,GATE} we concluded that
the shape of the cross-section very near threshold is indeed determined
by the three-body nature of the final state interaction.
Those calculations
were however made within a non-relativistic formalism,
with the only addition of non-relativistic kinematics for the pion.
Here we introduce and solve a relativistic formalism with the following features:

i) On one hand, covariant meson-nucleon amplitudes based on different
data analyses \cite{BAT1,BAT2,GRE1,GRE2,JULI}
of the coupled reactions
$\pi N \to \eta N$, $\eta N \to \eta N$ and
$\gamma  N \to \eta N$ are constructed for the first time.
The covariant meson-nucleon amplitudes are moreover conveniently boosted
(including their Dirac spin structure) to be embedded in the
meson production mechanism through which the reaction proceeds.
This mechanism is the meson-exchange
box diagram with the excitation of the $S_{11}$ resonance, 
represented in Fig. 1.
ii) On the other hand, for the calculation 
of the  $\eta d$ final state distortion,
a relativistic version of the 3-body equations is used, which incorporates 
relativistic kinematics and the boost of the two-body interactions. This
last effect for the meson-nucleon interactions is also studied.

There is a considerable dispersion
of the empirical values for the $\eta N$ scattering originated by different data analysis
\cite{BAT1,BAT2,GRE1,GRE2,JULI}.
The inclusion of relativity in the calculation of the
$np \rightarrow \eta d$ cross section is needed to
narrow the large uncertainty region for that scattering 
length. In the light of the relativistic description used here, the width of this region is seen
to be narrowed.

The next section describes the formalism: in Section 2.A
the relativistic meson-exchange driving term is introduced
and in Section 2.B the
three-body relativistic formalism for the  $\eta$d final
state interaction is addressed. In section III
the results are shown and discussed. 
Section IV summarizes the conclusions.

\section{Formalism}

\subsection{The Covariant $np \rightarrow \eta d$ Box Diagram}

We will start our discussion with the box diagram shown in Fig.1,
which together with the impulse term, has been considered as the basic meson production mechanism
\cite{LOCH,GAR2,GATE,JULI2,JULI3}.
\begin{figure}
\includegraphics[width=5.0cm,keepaspectratio]{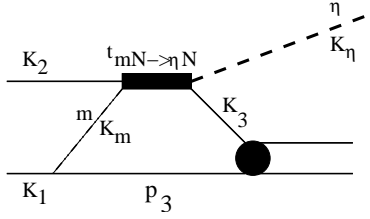} 
\caption{Meson-exchange mechanism for the reaction 
$np \rightarrow \eta d$
}
\label{FIG1}
\end{figure}

If one evaluates this Feynman diagram by putting the spectator nucleon on-the-mass-shell
one obtains
\begin{eqnarray}
A^{\mu_1 \mu_2}_{M_d}  =  \frac{1}{(2\pi)^3} \sum_{\mu} \int d\vec p_N&& 
\frac{M}{E_N}
 {{\bar{v}}^{\mu}}_N V_{dNN}^{M_d}
\frac{\not{p}_3+M}{p_3^2-M^2} t_{mN \rightarrow \eta N} {u}^{\mu_2}_2
\nonumber \\ 
& & \times \frac{1}{k_m^2-m_m^2}  {{\bar{u}}^{\mu}}_N  V_{mNN} 
{u}^{\mu_1}_1 ,
\label{eq1}
\end{eqnarray}


\noindent
where $V^{M_d}_{dNN}$, $V_{mNN}$ and $t_{mN \rightarrow \eta N}$ 
are respectively, the deuteron-nucleon-nucleon vertex, the $m$ meson-nucleon-nucleon
vertex, and the meson-nucleon $\rightarrow \eta$ nucleon t-matrix. The spinors
${u}^{\mu_1}_1$, ${u}^{\mu_2}_2$, ${u}^{\mu}_N$, correspond respectively the two initial and the intermediate
nucleon spinors,
and  ${\bar v}^{\mu}_N$ is the charged-conjugated spinor. The 3-momentum variables are defined
as shown on the diagram of Fig. 1.
 
One expects the effect of relativity to be important in the left hand-side of the
box diagram where the exchanged meson m is very far-off-the-mass-shell. 
However, on the right-on-side of the diagram the final $\eta$ is restricted
to the energy region of $E<100$MeV, so that the effects of relativity are not so crucial.
Therefore, we will write the propagator for the intermediate nucleon in terms
of positive and negative energy spinors 
and keep only the positive energy part:

\begin{eqnarray}
\frac{\not{p}_3 +M}{p_3^2-M^2} & = &
\frac{M}{E_3} \sum_{\mu_3} \frac{u^{\mu_3}_3(\vec p_3) {\bar u}^{\mu_3}_3(\vec p_3)}
{p_{03}-E_3} +
\frac{M}{E_3} \sum_{\mu_3} \frac{v^{\mu_3}_3(-\vec p_3) {\bar v}^{\mu_3}_3(-\vec p_3)}
{p_{03}+E_3}  
\nonumber \\ 
& & \rightarrow 
\frac{M}{E_3} \sum_{\mu_3} \frac{u^{\mu_3}_3(\vec p_3) {\bar u}_{\mu_3}^3(\vec p_3)}
{p_{03}-E_3}
\label{eqprop} 
\end{eqnarray}

If we now consider the deuteron wavefunction $\Psi^*_{M_d,\mu \mu_3}(\vec p\;)$ defined by the identification \cite{BUCK}
\begin{equation}
\frac{M}{E_3} \bar v^{\mu}_N(\vec k_3) V^{M_d}_{dNN} u_3^{\mu_3}(\vec p_3) \frac{1}{p_{03}-E_3} \equiv
\sqrt{2 \omega_d} (2 \pi)^{3/2} \Psi^*_{M_d,\mu \mu_3}(\vec p\;) 
\end{equation}
where $\vec p$ is the $NN$ relative momentum in the center of mass, we obtain from Eq.(\ref{eq1})
\begin{eqnarray}
A^{\mu_1 \mu_2}_{M_d} = \frac{1}{(2\pi)^{3/2}} \sum_{\mu,\mu_3} 
\int d\vec p_N&& \frac{M}{E_N}
\sqrt{2 \omega_d}\Psi^*_{M_d,\mu \mu_3}(\vec p\;)  {\bar u}_3^{\mu_3}(p_3)
 t_{mN \rightarrow \eta N} {u}^{\mu_2}_2 (p_2) 
 \nonumber \\
& & \times \frac{1}{p_m^2-m_m^2}  {{\bar{u}}^{\mu}}_N(p_N)  V_{mNN} 
{u}^{\mu_1}_1(p_1) .
\label{eq2}
\end{eqnarray}
In Eq. \ref{eq2} we made explicit the momentum dependence of the nucleon spinors.

The deuteron wavefunction is on the other hand calculated from
\begin{equation}
\Psi^*_{M_d,\mu \mu_3}(\vec p\;) = \sum_{L=0,2} \sum_{m_L, m_S}
C^{L \,\,\,\, \,\, \,1\,\,\, \,\, \,\, 1}_{m_L\,m_S\,M_d} \,\,\phi_L(p) 
Y_{L m_L}^*(\hat p)\,\, C^{1/2 \,\,\,1/2 \,\,\,1}_{\,\mu \,\,\,\,\, \,\mu_3 \,\, \,\,\,\, m_S},
\end{equation}
where $\phi_0(p)$ and $\phi_2(p)$ are the S- and D-wave components which we obtained
from the Paris potential.

For the $m$ meson-nucleon-nucleon vertex we take
\begin{equation}
V_{mNN}= g_m \gamma_5 f_m(k_m^2), \,\, \,   m= \pi, \eta
\end{equation}
and
\begin{equation}
V_{mNN}= g_m f_m(k_m^2), \,\, \,  m= \sigma
\end{equation}
where $k_m^2$ is the meson four-momentum squared and the form factor $f_m$ is chosen to have
the monopole form
\begin{equation}
f_m(k_m^2) = \frac{ \Lambda^2 -m_m^2}{\Lambda^2-k_m^2}
\end{equation}
with $\Lambda=1800$ MeV/$c$. 

Since the $\eta$ production near threshold is dominated by the $S_{11}$ resonance, the $m$
meson-nucleon $\rightarrow \eta$-nucleon $t$ transition operator is assumed to be generated
by a variable
mass isobar model consisting of a single isobar, the $S_{11}$. As in the framework
introduced in ref.\cite{GARZ} for the study of the pion induced eta production reaction,
the isobar model for meson-nucleon scattering used here is covariant,
and reads: 
\begin{eqnarray}
t_{mN \rightarrow \eta N} ({\vec p}^{\; 2},{\vec p}^{\; '2}, M_S) & = & \frac{(2 \pi)^2}{M}
\sqrt { {\omega}_m ({\vec p}^{\; 2}) {\omega}_{\eta} ({\vec p}^{\; '2}) E_N ({\vec p}^{\; 2})
 E_N({\vec p}^{\; '2})} 
 \nonumber \\
 & & \times  
 h_m(\vec p^{\; 2})\frac{\not{k_N}+\not{k_2}+M_S}{2M_S} h_{\eta}(\vec p^{\; '2})
\tau (M_S) 
\label{eqtmat}
\end{eqnarray}
\noindent
where $\omega_m ({\vec p}^{\; 2})= \sqrt{m_m^2+\vec p^{\; 2}}$, $E_N({\vec p}^{\; 2})=
\sqrt{M^2+{\vec p}^{\; 2}}$ are the on-shell energies respectively of the $m$ meson and of the nucleon
in the c.m. frame. Also one has
\begin{equation}
M_S=\sqrt{(k_m+k_2)^2}= \sqrt{(k_{\eta}+k_3)^2}.
\end{equation}
The meson-nucleon-isobar vertices are 
\begin{equation}
h_m({\vec p}^{\; 2}) = \sqrt {\frac{2M}{M+E_N(\vec p\;)} }
g_m({\vec p}^{\; 2}), \,\, \, m=\pi,\eta
\end{equation}
and
\begin{equation}
h_m({\vec p}^{\; 2}) =\frac{M}{\sqrt{\vec p^{\; 2}}}\sqrt {\frac{2M}{M+E_N(\vec p\;)} }
g_{m}({\vec p}^{\; 2}) \gamma_5, \,\, \, m=\sigma
\end{equation}
Here, three-momentum squared $\vec p^{\; 2}$ ($\vec p^{\; '2}$) is the meson-nucleon
relative initial (final) three-momentum in the c.m frame. In particular, it is
related to Lorentz invariant quantities as
\begin{equation}
\vec p^{\; 2}=\frac{(M_S^2+M^2-k_m^2)^2}{4 M_S^2} - M^2.
\end{equation}
The isobar propagator $\tau(M_S)$  is obtained from a separable potential model
describing the coupled $\eta N-\pi N-\sigma N$ two-body subsystem. The
corresponding two-body $t-$matrix (\ref{eqtmat}) is also separable in any reference frame.
We note that the $\sigma N$ channel stands for the $\pi\pi N$ inelasticity. Also,
in variance with refs. \cite{WILKIN} we do not consider $\rho$-exchange. Recently in ref. \cite{JULI2}
it is shown that the exact numerical treatment of the initial state interaction reduces
significantly the $\rho$-exchange diagram, relatively to the other meson exchanges.
 
The matrix element of the transition operator on Eq. (\ref{eqtmat})
in the 2 body meson-nucleon c.m. system, 
where ${\vec k}_m+{\vec k}_2=\vec k_{\eta}+\vec k_3=0$,
in the basis states of the nucleon spinors is given by
\begin{eqnarray}
{{\bar{u}}^{\mu_3}}_3 t_{mN \rightarrow \eta N} {u}^{\mu_2}_2 & = &
\frac{2\pi}{M} \delta_{\mu_2\mu_3}
\sqrt { {\omega}_m ({\vec p}^{\; 2}) {\omega}_{\eta} ({\vec p}^{\; '2}) E_N ({\vec p}^{\; 2})
 E_N({\vec p}^{\; '2})}
 \nonumber \\
& & \times g_{m}({\vec p}^{\; 2}) \tau(M_S) g_{\eta}({\vec p}^{\; '2}).
\label{eqrelt}
\end{eqnarray}

We consider nucleon 2 with momentum $\vec q_{N}$ in the positive direction 
of the z-axis and the $\eta$ meson 3-momentum in the $xz$ plane with polar angle $\theta$. Then
the amplitude $\cal{A}$ of Eq. (\ref{eq2}) satisfies the symmetry property
\begin{equation}
A_M^{\mu_1 \mu_2} (\vec q_{N}, \theta) = -(-1)^{M+\mu_1+\mu_2} 
A_M^{\mu_1 \mu_2}  (-\vec q_{N}, \pi-\theta).
\end{equation}

In the isospin formalism the neutron and proton are identical particles, so that
the initial $np$ state must be antisymmetrized under the exchange of nucleons 1 and
2. However, the system is in a pure total isospin $0$ state, which means that 
the initial $np$ state must be symmetric under the exchange of space and spin variables.
Therefore, the correctly antisymmetrized amplitude $\bar A_M^{\mu_1 \mu_2}$
for the $np \rightarrow \eta d$ process is 
\begin{eqnarray}
\bar A_M^{\mu_1 \mu_2} & = &\frac{1}{\sqrt{2}} \left[A_M^{\mu_1 \mu_2} 
(\vec q_{N},\theta) + A_M^{\mu_2 \mu_1}(-\vec q_{N},\theta) \right]
\nonumber \\
& = &\frac{1}{\sqrt{2}} \left[A_M^{\mu_1 \mu_2} (\vec q_{N},\theta) - (-1)^{M+\mu_1+\mu_2} A_M^{\mu_2 \mu_1}(\vec q_{N},\pi-\theta) \right].
\end{eqnarray}

\subsection{The $\eta d$ scattering and the boost of the two-body interactions}
In our previous work \cite{GATE} we presented the formalism of $\eta d$
elastic scattering based on nonrelativistic Faddeev equations. 
Since we are now discussing relativistic effects in the $np\to\eta d$ process,
it becomes necessary to perform here also a relativistic calculation
of the $\eta d$ elastic channel responsible for the
final-state interaction in the $np\to\eta d$ reaction. To generate
the necessary $\eta d$ distorted waves, we apply to the $\eta d$
elastic channel the relativistic formalism in momentum space presented in Ref. \cite{GARX}. This formalism
generalizes in a straightforward way the non-relativistic Faddeev equations.
It incorporates relativistic kinematics and, importantly, also the boosts of the two-body interactions
to the three-body c.m. frame


Firstly, the main feature of the formalism
of \cite{GARX} is to consist of a set of relativistic but 3-dimensional integral Faddeev-type equations
obtained from a field theory in which the three particles are kept on their mass shells 
in all intermediate
states. Accordingly, in what follows the quantity $k_i$ does not refer to the 
four-momentum of particle $i$, but to the magnitude of the 3-momentum $\vec k_i$. Secondly,
in order to transform correctly all physical quantities from the two-body to the three-body reference
frames, and after considering the energy conservation constraint,
one writes the invariant momentum space volume element for the three particles in terms of
the two relative Jacobi variables $\vec p_i$ and $\vec q_i$, 
\begin{eqnarray}
d \cal{V} & = & {d\vec k_1\over 2\omega_1(k_1)}
 {d\vec k_2\over 2\omega_2(k_2)}
 {d\vec k_3\over 2\omega_3(k_3)}\delta(\vec k_1+\vec k_2+\vec k_3)
\nonumber \\ & &
= {\omega(p_i)\over 8W_i(p_iq_i)\omega_i(q_i)\omega_j(p_i)\omega_k(p_i)}
d\vec p_i d\vec q_i,
\label{eq3}
\end{eqnarray}
The variable $\vec p_i$ is the relative momentum of the pair $jk$ measured in 
the c.m. frame of the pair (that is, the frame in which particle $j$
has momentum $\vec p_i$ and particle $k$ has momentum $-\vec p_i$), and
$\vec q_i=-\vec k_i$ is the relative momentum between the pair $jk$ and
the spectator particle $i$, measured in the c.m. frame of the three particles,
(in which the pair $jk$ has total momentum $\vec q_i$ and particle $i$ 
has momentum $-\vec q_i$). The energy of the $jk$ pair in its c.m. frame is
\begin{equation}
\omega(p_i)=\sqrt{m_j^2+p_i^2}+\sqrt{m_k^2+p_i^2},
\label{eq4}
\end{equation}
the total energy of the pair is
\begin{equation}
W_i(p_iq_i)=\sqrt{\omega^2(p_i)+q_i^2},
\label{eq5}
\end{equation}
and the invariant energy of the three particles is written
\begin{equation}
W(p_iq_i)=\omega_i(q_i)+W_i(p_iq_i).
\label{eq6}
\end{equation}
Equations (\ref{eq3}) throughout (\ref{eq6}) determine the transformation of the matrix elements of the two-body
potential $V$ from the two-body to the three-body reference frames as in
\cite{GARX}
\begin{eqnarray}
<\vec p_i\vec q_i|V|\vec p_i^{\; \prime}\vec q_i^{\; \prime}> & = & \left[
 {W_i(p_iq_i)\omega_j(p_i)\omega_k(p_i)W_i(p_i^\prime q_i)\omega_j(p_i^\prime)
\omega_k(p_i^\prime)\over \omega(p_i)\omega(p_i^\prime)}\right]^{1/2}
\nonumber \\ & &
\times 8\omega_i(q_i)\delta(\vec q_i-\vec q_i^{\; \prime})
V(\vec p_i,\vec p_i^{\; \prime}),
\label{eq7}
\end{eqnarray}
which in turn defines the boosted matrix elements of the two-body t-matrix.
These are given by
\begin{eqnarray}
<\vec p_i\vec q_i|t|\vec p_i^{\; \prime}\vec q_i^{\; \prime}> & = & \left[
 {W_i(p_iq_i)\omega_j(p_i)\omega_k(p_i)W_i(p_i^\prime q_i)\omega_j(p_i^\prime)
\omega_k(p_i^\prime)\over \omega(p_i)\omega(p_i^\prime)}\right]^{1/2}
\nonumber \\ & &
\times 8\omega_i(q_i)\delta(\vec q_i-\vec q_i^{\; \prime})
t(\vec p_i,\vec p_i^{\; \prime};q_i),
\label{eq8}
\end{eqnarray}
where $t(\vec p_i,\vec p_i^{\; \prime};q_i)$ satisfies the Lippmann-Schwinger
equation with a propagator corresponding to relativistic kinematics defined
by Eq. (\ref{eq6}):
\begin{eqnarray}
t(\vec p_i,\vec p_i^{\; \prime};q_i) & = &
V(\vec p_i,\vec p_i^{\; \prime})+\int d\vec p_i^{\; \prime\prime}
V(\vec p_i,\vec p_i^{\; \prime\prime})
\nonumber \\ & &
\times {1\over W_0-W(p_i^{\prime\prime}q_i)+i\epsilon}
t(\vec p_i^{\; \prime\prime},\vec p_i^{\; \prime};q_i).
\label{eq9}
\end{eqnarray}
The variable $W_0$ is the invariant energy of the system. For only S-wave 
two-body interactions Eq. (\ref{eq9}) becomes
\begin{eqnarray}
t(p_i,p_i^\prime;q_i) & = &
V(p_i,p_i^\prime)+\int_0^\infty {p_i^{\prime\prime}}^2
d p_i^{\prime\prime}
V(p_i,p_i^{\prime\prime})
\nonumber \\ & &
\times {1\over W_0-W(p_i^{\prime\prime}q_i)+i\epsilon}
t(p_i^{\prime\prime},p_i^\prime;q_i).
\label{eq10}
\end{eqnarray}

In the particular case of the coupled $\eta N-\pi N-\sigma N$ subsystem 
(we
take $m_\sigma=2m_\pi$, since the $\sigma N$ channel simulates the $\pi\pi N$ inelasticity),
these three different meson-nucleon channels are connected among each other 
through the $S_{11}$ partial wave. For each transition, we use rank-one separable potentials
of the form
\begin{equation}
V_{mm'}(p_i,p_i^\prime)=-g_m(p_i)g_{m'}(p_i^\prime);\,\,\,\,\,\,\,\,\,\,
(m,m'=\eta,\pi,\sigma)
\label{eq11}
\end{equation}
where the functions $g_m$ are as in ref. \cite{GATE},
\begin{equation}
g_m(p_i)=\sqrt{\lambda_m}{A_m+p_i^2\over (\alpha_m^2+p_i^2)^2};
\,\,\,\,\,\,\,\,\,\,(m=\eta,\pi)
\label{eq12}
\end{equation}
\begin{equation}
g_m(p_i)=\sqrt{\lambda_m}{p_i\over (\alpha_m^2+p_i^2)^2};
\,\,\,\,\,\,\,\,\,\,(m=\sigma)
\label{eq13}
\end{equation}
so that the solution of Eq. (\ref{eq10}) is
\begin{equation}
t_{mm'}(p_i,p_i^\prime;q_i)=g_m(p_i)\tau(q_i)g_{m'}(p_i^\prime),
\label{eq14}
\end{equation}
with
\begin{equation}
\tau^{-1}(W_0,q_i)=-1-\sum_{m=\eta,\pi,\sigma}\int_0^\infty p_i^2 dp_i
{g_m^2(p_i)\over W_0-W(p_i q_i)+i\epsilon}.
\label{eq15}
\end{equation}

We will now make the connection between the boosted two-body t-matrix
elements
and the ones introduced in the previous section. For that, we take
Eq. (\ref{eq8})  in the two-body c.m. frame, where $q_i=0$, and obtain
\begin{eqnarray}
<\vec p_i\vec 0|t|\vec p_i^{\; \prime}\vec q_i^{\; \prime}> & = & \left[
 {\omega_j(p_i)\omega_k(p_i)\omega_j(p_i^\prime)
\omega_k(p_i^\prime)}\right]^{1/2}
\nonumber \\ & &
\times 8m_i\delta(\vec 0-\vec q_i^{\; \prime})
t(\vec p_i,\vec p_i^{\; \prime};0),
\label{eq16}
\end{eqnarray}
We verify this way that the t-matrix given by Eq.(\ref{eq16}), and  generated by our separable 
potential model on Eq.(\ref{eq11}), is proportional to Eq.(\ref{eqrelt}).
The different multiplicative factors
come from normalization conventions for the two body 
momentum basis states, taken differently in relativistic and non-relativistic Faddeev-type 
formalisms.

The driving terms of 
the Faddeev equations for $\eta d$ elastic scattering given by Eqs.
(20)-(22) of Ref. \cite{GATE} are here  modified by the inclusion of
relativistic kinematics and an invariant three-body volume element
by making the replacement
\begin{eqnarray}
{1\over E-p_j^2/2\mu_j-q_j^2/2\nu_j+i\epsilon} & \to & \left[
 {W_i(p_i^\prime q_i)\omega_j(p_i^\prime)\omega_k(p_i^\prime)
W_j(p_jq_j)\omega_k(p_j)
\omega_i(p_j)\over \omega(p_i^\prime)\omega(p_j)\omega_i(q_i)\omega_j(q_j)
}\right]^{1/2}
\nonumber \\ & &
\times {1\over \omega_k(q_k)}{1\over W_0-W(p_jq_j)+i\epsilon}.
\label{eq18}
\end{eqnarray}
with $p_i^\prime$, $p_j$, and $\omega_k(q_k)$ defined by Eqs. (70), (71),
and (66) of Ref. \cite{GARX}.

In the case of the $NN$ interaction for the intermediate re-sacttering series given
in \cite{GATE} by eqs. (19) and (20), and represented therein on Fig. 2, we used in \cite{GATE} the PEST
separable model of Ref.\cite{PEST}, which is based in the nonrelativistic
Lippmann-Schwinger equation. Here, in order to make this interaction 
consistent with the relativistic Lippmann-Schwinger equation given by eq. (\ref{eq10})
we re-adjusted numericaly the strength of the potential such that the deuteron pole
appears at the right position. This amounts to multiplying the original PEST
potential of Ref. \cite{PEST} by
the factor 0.78805.
As for the $\eta NN$ 
coupling constant, it was given the reasonable
value \cite{BONN,NIJM,BENM,TIAT,ZHU} $g_\eta^2/4\pi$ = 1 
while for the $\sigma NN$ coupling constant we used the 
value of Ref. \cite{HOLIN} $g_\sigma^2/4\pi$ = 8.

\section{Results}

We give in table I the parameters of the seven models corresponding to
the description of
the meson-nucleon amplitude analyses of
Refs. \cite{BAT1,BAT2,GRE1,GRE2,JULI}
by eqs. (\ref{eq10})-(\ref{eq14}). The agreement
of the amplitudes obtained with those analyses is at least as good  
as the ones shown in Fig. 1 in ref. \cite{GATE}, and therefore we
do not show here 
the corresponding figure.

In all, to the exception of model 0,
the parameters for $\sigma$-exchange
are the ones that deviate less from the
parameters obtained in ref. \cite{GATE} from a non-relativistic
calculation.
However, due to its small mass, the pion contribution
is affected by the relativistic treatment.
In models 2-6 relativity increases slightly the momentum
range parameter $\alpha_{\pi}$. However,
the strength parameter $A_{\pi}$
increases also. Since, as seen from Eq.(\ref{eq12}),
it defines the
weight of the small versus the large momentum region,
this way it compensates the extra weight of the high momentum tail
originated by the increase of the range.
As for 
the $\eta$N channel, the changes in the two parameters also
balance out.
\begin{figure}
\epsfig{file=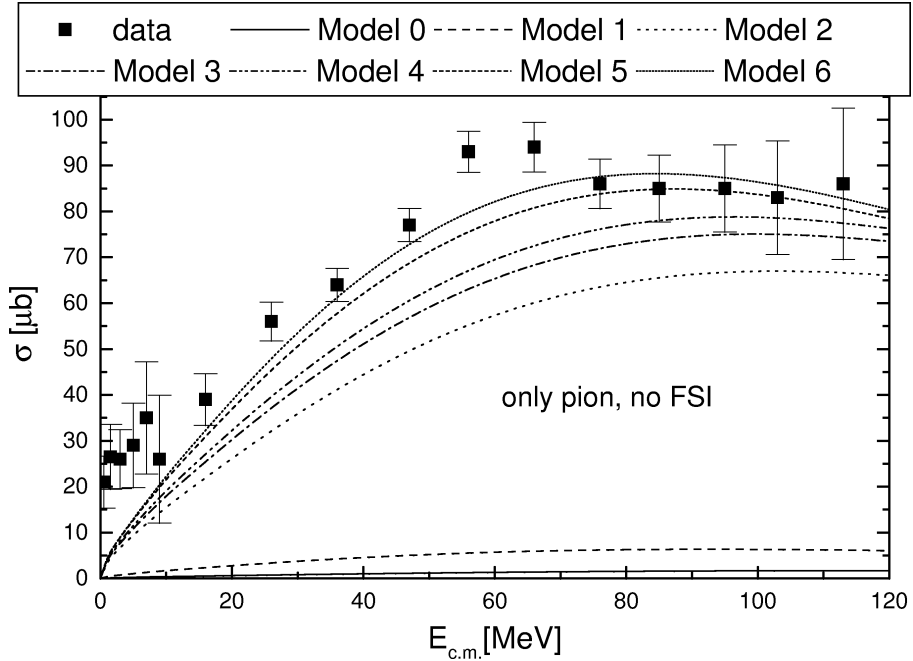,width=12.0cm} 
\caption{Total cross section of the
reaction $np \rightarrow \eta d$ for the seven relativistic
meson-nucleon interaction models considered. Only pion
exchange is considered in the box diagram represented by Fig. 1.
No $\eta$d final state interaction included. Data from refs. \cite
{CAL1} and \cite{CAL2}.}
\label{FIG2}
\end{figure}

Models 1, and specially model 0, which correspond to
smaller values for the
$\eta$N scattering lenghts, behave differently than the others,
since the pion range is seen to decrease.
Moreover, for model 0 the pion low-momentum strength $A_{\pi}$ uniquely decreases
more than the range parameter $\alpha_{\pi}$. As a net result the weigth of the high
momentum range versus the low momentum range is increased.
This relative
weight of small and large momenta reflects then on the behavior 
of the cross-section for $np \rightarrow \eta d$, as we will see below.
The difference in behavior of model 0 may be due to the
inclusion of $\rho$-exchange
in the meson-nucleon amplitude as explained in ref. \cite{JULI2},
which has the role of
counteracting the high momenta contribution of pion exchange.

We will turn now to the results obtained for the $\eta d$
three-body elastic channel.
We give in table II the predictions for
the three-body $\eta d$ scattering length obtained
using the seven two-body 
$\eta N$-$\pi N$-$\sigma N$ coupled interaction models.
We present the results corresponding to
the box diagram or driving term of the 
Faddeev equations (see Fig. 2 of Ref. \cite{GATE})
including the different meson-exchanges,
one by one, in successive cumulative steps: only
$\eta$ exchange, $\eta+\pi$ exchange or $\eta+\pi+\sigma$ exchange.
To conclude about the extent of the relativistic effects
we show for each  model the non-relativistic
results  of \cite{GATE} (lines labeled "NR").
The effect of the relativistic treatment
on the three-body scattering length
can be seen 
clearly in the contribution
of the pion, which is the lightest meson.  
Its contribution to the 3 body scattering length is negligible in all
non-relativistic models \cite{GATE}.  
Compared with the corresponding results of \cite{GATE}, 
the relativistic results are still quite similar 
to those of the nonrelativistic case, to the exception
of model 0. 
The pion exchange contribution to the scattering length is now
quite important in the case of this model. This happens because
in model 0, contrarily to the other models, 
the relativistic changes in the range $\alpha_{\pi}$
and low-momentum strength $A_{\pi}$ parameters are not balanced,
as in the other models.

Besides the changes that the relativistic two-body models induce
in the pion exchange contribution, 
the comparison done on table II gives also indirect
information on the
magnitude of the boosts of the two body
interactions within the 3-body system. 
Given the agreement observed in most cases
between the relativistic and non-relativistic models,
boost effects do not appear relevant.
Since the energies involved ($\sim$ 100 MeV), are
much smaller than the masses of the $\eta$ and the nucleon,
this is expected.

\begin{figure}
\epsfig{file=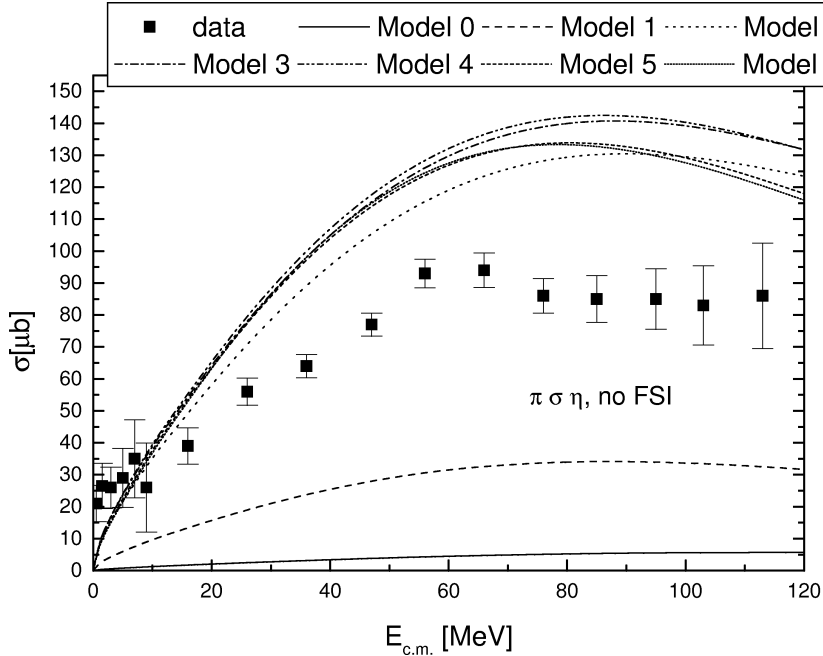,width=12.0cm} 
\caption{The same as Fig.2 but with inclusion of 
$\pi$, $\eta$ and $\sigma$ exchanges in the diagram
represented by Fig.1.}
\label{FIG3}
\end{figure}

We show in Fig. 2 the cross section of the $np \rightarrow \eta d$ 
process when only pion exchange is included in the box diagram
(see Fig. 1) and {\bf no} final-state interaction is included. We considered
a reduction factor of 5 corresponding to the
initial-state interaction  \cite{GATE}. Although some of the models
(those with a large $\eta N$ scattering length) predict more or less
the right magnitude for the cross section at large energies, near
threshold they all fail to reproduce the enhancement shown by the data.
We notice that the models with larger relative low-momentum
strength parameter
$A_{\pi}$, larger absolute pion strength $\lambda_{\pi}$ and
smaller absolute $\eta$ strength $\lambda_{\eta}$ ,
are closer to the data away from threshold, where indeed only
small momentum transfer is needed.
 
We show in Fig. 3 the corresponding results when in 
addition the contribution of the exchanges of the $\eta$ and
$\sigma$ mesons are included in the box diagram. 
From both  Figs. 2 and 3 it 
is clear that the dominant exchange mechanism
for  the $np\to\eta d$ process 
is pion exchange. This conclusion was also found by
the substantially different
calculations of refs. \cite{JULI2,NAKAYAMA}.

We consider next the situation with regard to the  final $\eta d$
distortion
of
the $np\to\eta d$ process.
We show in Fig. 4 the results when in addition one includes the 
final-state interaction. The models with a large $\eta N$ scattering
length give a good description of the data near threshold.
This was already the case for the non-relativistic case in ref.
\cite{GATE} (see fig. 6 therein). The new feature of the relativistic
calculation is that  
the high-energy end is now described by those models. However,
they fail to reproduce the shape of the cross 
section in the intermediate region. The good description of the cross
section at the high
energy end by models 2-6 is due to the modification of the range and strength parameters
for the pion in the dynamical two-body models
- and does not happen for models 0 and 1.

\begin{figure}
\epsfig{file=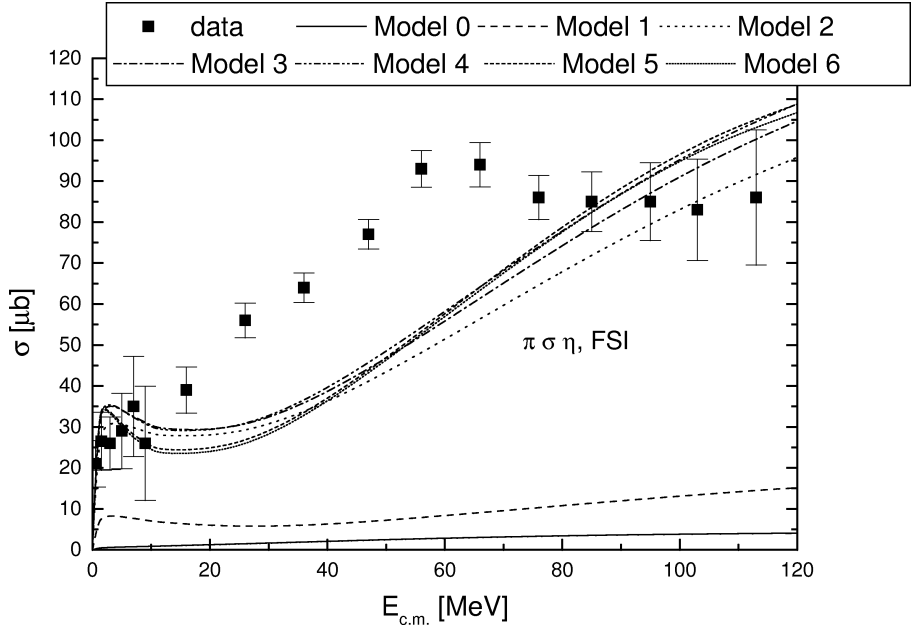,width=12.0cm} 
\caption{The same as Fig.3 but with the inclusion of 
$\eta$d final state interaction.}
\label{FIG4}
\end{figure}

Finally, we introduced in the box diagram the contribution of the
$\eta^\prime$ meson with coupling constants adjusted so as to reproduce the
cross section near threshold. Since this is a heavy meson exchange,
this process acts like a background correction to
the isobar model of the nucleon-meson amplitude.
We show the results in Fig. 5. This figure is indicative
that a reasonable description of the data could be obtained
with a model in between model 0 and model 1, i.e., with a
$\eta N$ scattering length larger than 0.42 fm and smaller than
0.72 fm. A finer tuning beyond this point demands an exact treatment of
the initial state reduction effect, which as explained in ref \cite{JULI2},
may be different for the different meson exchanges.

\begin{figure}
\epsfig{file=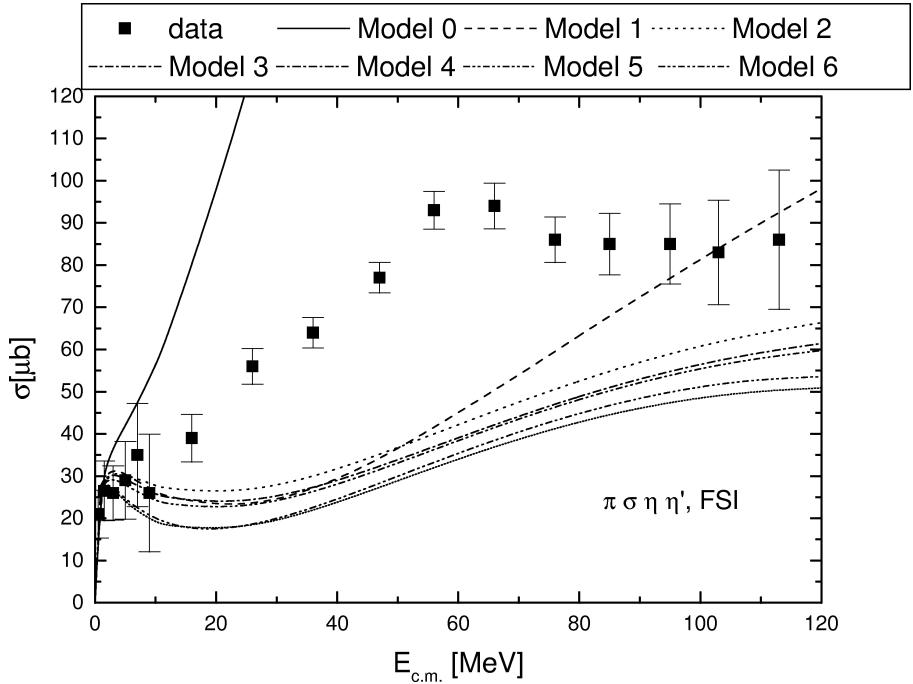,width=12.0cm} 
\caption{The same as Fig. 4 but with the inclusion of 
$\eta$' exchange in the box diagram represented by Fig. 1.}
\label{FIG5}
\end{figure}

\section{Conclusions}

To summarize, a relativistic calculation of the $np \rightarrow \eta d$,
based on 
meson exchange production mechanisms,
confirms that the final state $\eta$d interactions is 
very important near threshold,
as found before in non-relativistic models. This interaction
alone explains the enhancement effect observed in 
the cross section near threshold.

Although the pion exchange visibly does not describe the data
near threshold,
an important conclusion is that
the relativistic
pion exchange contribution dominates the reaction
exchange mechanisms. Moreover, it describes 
the high energy end of
the measured cross-section, which was not the case for the
non-relativistic version.

Finally, relativistic models corresponding to relatively lower 
values of the $\eta$d scattering length,
e.g. the Julich model, when compared to their non-relativistic
counterparts, have the pion strength in
the large momentum tail accentuated relatively to the small momentum
region.  Simultaneously, however, the overall pion strength is
reduced.
These two features  seem to be needed to describe successfully the energy dependence
of the cross-section,
in the threshold region, as well as away from the 
threshold energy region. 

The control of relativistic effects are therefore important in 
phenomenological analysis of the reaction. Namely, it
narrows considerably the uncertainty in the knowledge of the
$\eta$N scattering length. A value for this one 
between 0.42 fm and
0.72 fm seems to be indicated by this study.

\newpage
This work was supported in part by 
COFAA-IPN (M\'exico) and
by Funda\c c\~ao para a Ci\^encia e a Tecnologia,
under contract POCTI/FNU/50358/2002.



\newpage

\begin{table}
\caption{Parameters of the $\eta N$-$\pi N$ separable potential 
models fitted to the $S_{11}$ resonant amplitudes given in
Refs. [7-10].}
\begin{tabular}{ccccccccccc}
 Model & Ref. & $a_{\eta N}$& $\alpha_{\eta}$  & $A_\eta$ &
$\lambda_{\eta}$  & $\alpha_\pi$  
& $A_\pi$  & $\lambda_\pi$ & $\alpha_\sigma$ & $\lambda_\sigma$    \\
\hline
\\
 0  & [10] & 0.42+i0.34 
 &            5.85798
 &          7.20057
 &         -202.573
 &           0.384338
 &           0.00306689
 &           -0.0804278
 &           0.808
 &          -0.155061\\
 1 &[7] & 0.72+i0.26 
 &          29.9983
 &         359.211
 &       -5991.79
 &           2.28053
 &           1.08638
 &          -0.0660518
 &           8.0
 &         -239.860 \\
 2 &[8]  & 0.75 + i0.27 
 &           6.80695
 &        409.632
 &           -0.0735564
 &            9.17614
 &           1.46599
 &        -701.087
 &            8.0
 &        -816.460 \\
 3 &[9](D)  &  0.83 + i0.27 
 &          5.43840
 &          74.9154
 &           -0.387884
 &            8.83448
 &           0.449176
 &        -654.504
 &           8.0
 &        -760.560 \\
 4 &[9](A)  &  0.87+i0.27 
 &            4.35990
 &           30.3941
 &          -0.376959
 &           8.96712
 &           0.270940
 &         -687.477
 &           8.0
 &        -618.431 \\
 5 &[9](B)  & 1.05 + i0.27 
 &           2.04950
 &           2.60222
 &           -0.102332
 &            9.71806
 &           0.192626
 &         -849.271
 &           8.0
 &         -236.559 \\
 6 &[9](C)  & 1.07 + i0.26 
 &           1.99979
 &           2.28184
 &          -0.105698
 &           9.76374
 &           0.0702236
 &         -861.215
 &           8.0
 &         -174.670 \\
\end{tabular}
\end{table}

\begin{table}
\caption{$\eta d$ scattering length (in fm) predicted by the seven separable
potential models of the coupled $\eta N$-$\pi N$-$\sigma N$ system. 
We give the results obtained including only
$\eta$-exchange, $\eta$- and $\pi$-exchange, and $\eta$-$\pi$- and
$\sigma$-exchange in the driving terms. Comparison with results of ref.
\cite{GATE} is provided on the lines labeled NR.}
\begin{tabular}{ccccccc}
 & Model & $a_{\eta N}$ &  $\eta$  & $\eta +\pi$  
& $\eta +\pi +\sigma$  &  \\
\hline
  & 0    & 0.42+i0.34 & 0.88+i1.34 & 0.39+i1.67 & 0.23+i1.68 & \\
NR&  0    & 0.42+i0.34 & 1.01+i1.24 & 1.00+i1.28 & 0.99+i1.28 & \\
\hline
 & 1    & 0.72+i0.26 & 2.59+i1.84 & 2.67+i1.90 & 2.67+i1.90 &  \\
NR &1    & 0.72+i0.26 & 2.53+i1.51 & 2.56+i1.51 & 2.57+i1.51 &  \\
\hline
  & 2    & 0.75+i0.27 & 2.73+i1.66 & 2.78+i1.68 & 2.78+i1.68 &  \\
NR &2    & 0.75+i0.27 & 2.75+i1.64 & 2.75+i1.62 & 2.76+i1.62 &  \\
\hline
  &3    & 0.83+i0.27 & 3.23+i1.88 & 3.28+i1.91 & 3.29+i1.91 &  \\
NR&  3    & 0.83+i0.27 & 3.28+i1.93& 3.28+i1.91 & 3.30+i1.91 &  \\
\hline
 &4    & 0.87+i0.27 & 3.45+i1.92 & 3.50+i1.95 & 3.51+i1.95 &  \\
NR& 4    & 0.87+i0.27 & 3.55+i2.07& 3.56+i2.05 & 3.57+i2.04 &  \\
\hline
  &5    & 1.05+i0.27 & 4.72+i2.47 & 4.80+i2.52 & 4.80+i2.52 &  \\
NR& 5    & 1.05+i0.27 & 4.91+i2.72 & 4.92+i2.70 & 4.93+i2.70 &  \\
\hline
 &6    & 1.07+i0.26 & 5.00+i2.54 & 5.09+i2.60 & 5.09+i2.60 &  \\
NR&6    & 1.07+i0.26 & 4.77+i2.25& 4.79+i2.25 & 4.79+i2.24 &  \\
\end{tabular}
\end{table}

\end{document}